\documentclass[aps,prd,preprint,groupedaddress,showpacs]{revtex4-1}

\usepackage{amsmath}
\usepackage{amssymb}
\usepackage{epsfig}
\usepackage{graphicx}
\usepackage{braket}
\newcommand{\dd}{\mathrm{d}}
\newcommand{\ii}{\mathrm{i}}

\begin{document}

\preprint{}

\title{\boldmath Free compact boson on branched covering of the torus}


\author{Feihu Liu}
\email[]{liufeihu04@hotmail.com}
\affiliation{University of Electronic Science and Technology of China,\\North Jianshe Road, Chengdu, Sichuan, P.R.China.}


\date{\today}

\begin{abstract}
We have studied the partition function of a free compact boson on a $n$-sheeted covering of torus gluing along $m$ branch cuts. It is interesting because when the branched cuts are chosen to be real, the partition function is related to the $n$-th R\'enyi entanglement entropy of $m$ disjoint intervals in a finite system at finite temperature.  After proposing a canonical homology basis and its dual basis of the covering surface, we find that the partition function can be written in terms of theta functions.
\end{abstract}


\maketitle

\section{Introduction}
\label{sec:intro}
Conformal field theory on higher genus Riemann surfaces is a fruitful field for both physics and mathematics. It has many important applications in physics, for example, string perturbation theory and statistical physics. Recently, entanglement entropy of two dimension field theory has attracted much attention due to its calculability. In Ref.\cite{ee-and-cft,Calabrese:2009ez}, the R\'enyi entanglement entropy of a free complex boson compactified on a torus was proved to be the same as the partition function for the $n$-sheeted covering surfaces of $\mathbb{CP}^1$. Indeed, there are three different cases: 1). The subsystem is in a infinite system at zero temperature; 2). The subsystem is in a finite system at zero temperature; 3). The subsystem is in a infinite system at finite temperature.
The R\'enyi entanglement entropy for these three cases are related by a conformal transformation and all can be derived from the partition function on the $n$-sheeted covering of $\mathbb{CP}^1$. More specifically, for $n$-th R\'enyi entanglement entropy of $m$ disjoint intervals, the corresponding surface should be represented by a singular $Z_n$ curve
\begin{equation}\label{eq:zncurve}
	{y}^n\equiv \prod_{i=0}^{m-1}(z-z_{2i-1})\prod_{j=1}^{m}(z-z_{2j})^{n-1}.
\end{equation}
This curve has been well studied in Ref.\cite{enolski:2006,Bobenko:2011}, where the period matrix and the Thomae formula are given explicitly. As a matter of fact, in Ref.\cite{Coser:2013qda}, the partition function of free compact boson on surface \eqref{eq:zncurve} has been derived by using the results of Ref.\cite{Bobenko:2011}.

For $n$-th R\'enyi entanglement entropy of $m$ disjoint intervals in a \emph{finite} system at \emph{finite} temperature, the corresponding Riemann surface should be the branched covering of torus, which is the main studying object of this paper.
These surfaces are constructed by gluing $n$ replica torus along $m$ branch cuts denoted by $\mathcal{T}_{n,m}$. In order to find the partition function, one can introduce the so-called twist fields, which first appear in the orbifold theory \cite{Dixon:1986qv,Dijkgraaf:1989hb,Bershadsky:1986fv,Knizhnik:1987xp}. By inserting a pairs of twist-antitwist fields at the ends of each interval, the total partition function on $\mathcal{T}_{n,m}$ is just the correlation function of twist fields on the torus \cite{ee-and-cft}
\begin{equation}
	\label{eq:totalcorrelationfunction}
	\begin{split}
		Z&=\prod_{k=0}^{n-1}\braket{\sigma_{k}(z_1,\bar{z}_1)\sigma_{n-k}(z_2,\bar{z}_2)\cdots\sigma_{k}(z_{2m-1},\bar{z}_{2m-1})\sigma_{n-k}(z_{2m},\bar{z}_{2m})}\\
		&=\sum_{windings}\prod_{k=0}^{n-1}Z_{qu}(k)e^{-S_{cl}(k)}.
	\end{split}
\end{equation}
Here $\sigma_{k}(z_i,\bar{z}_i)$ is the twist field which introduces the local monodromy around point $z_i$: 
\begin{equation}
	\begin{split}
		\partial\phi\rightarrow e^{2\pi \ii k/n}\partial\phi,\quad
		\partial\bar{\phi}\rightarrow e^{-2\pi \ii k/n}\partial\bar{ \phi},
	\end{split}
\end{equation}
and $\sigma_{n-k}(z_i,\bar{z}_i)$ denotes the antitwist field.
Note that the quantum part and the classical part of the correlation function have been separated for each $k$-mode. However the summation must be performed after the product, because the compactification condition introduces non-trivial couplings between the winding numbers for different $k$-modes, which is actually an obstacle to generalizing the results to higher genus cases.

By far, the most well studied example is the R\'enyi entropy for a single interval at finite temperature, which is corresponding to the surface $\mathcal{T}_{n,1}$.
One can find the calculation for free boson theory in Ref.\cite{Datta:2013hba,Chen:2015cna}. While in Ref.\cite{Schnitzer:2015ira,Schnitzer:2016xaj}, the R\'enyi entropy was obtained by the similar way. 

The main purpose of this paper is to evaluate \eqref{eq:totalcorrelationfunction} for free compact boson defined on $\mathcal{T}_{n,m}$.
Unlike the singular $Z_n$ curve \eqref{eq:zncurve}, the basis of holomorphic differentials and the homology basis of $\mathcal{T}_{n,m}$ are not known. However, by using the cut abelian differentials defined in Ref.\cite{Atick:1987kd}, we are able to construct the basis explicitly. In this way, we do \emph{not} need to separate the classical part into different $k$ modes, therefore the winding numbers are all independent and there will be no annoying infinity that appears in the summation.

\section{Quantum partition function on $\mathcal{T}_{n,m}$}

Since the quantum part of the partition function doesn't depend on the windings, so the correlation function can be derived following Ref.\cite{Atick:1987kd}. 
Let's first define the two Greens functions
\begin{equation}
	\label{eq:greensfunctions}
	\begin{split}
		g(z,\omega)&=\frac{\braket{\partial \phi(z)\partial \bar{\phi}(\omega)\prod_{i=1}^{2m}\sigma_{k_i}(z_i,\bar{z}_i)}}{\braket{\prod_{i=1}^{2m}\sigma_{k_i}(z_i,\bar{z}_i)}},\\
		h(\bar z,\omega)&=\frac{\braket{\bar{\partial} \phi(\bar{z})\partial \bar{\phi}(\omega)\prod_{i=1}^{2m}\sigma_{k_i}(z_i,\bar{z}_i)}}{\braket{\prod_{i=1}^{2m}\sigma_{k_i}(z_i,\bar{z}_i)}}.
	\end{split}
\end{equation}
The Greens function $g(z,\omega)$ is a doubly periodic holomorphic function for both $z$ and $\omega$. 
Based on the OPE
\begin{equation}\label{eq:ope}
	\partial \phi(z) \partial \phi(\bar{\omega})\sim \frac{1}{(z-\omega)^2}+T(\omega),,
\end{equation}
we know that $g(z,\omega)$ should have a double pole with coefficient $1$ as $z\rightarrow \omega$ but no simple pole. Additionally, $g(z,\omega)$ should satisfy the correct local monodromy for $z$ and $\omega$ around the branch points. It is similar for $h(\bar{z},\omega)$, the only difference is that there are no poles as $\bar{z}\rightarrow \omega$.
It turns out that these conditions are quite restrict. The two Greens functions can be fixed up to an undetermined function which is irrelevant to the partition function.
As proposed in Ref.\cite{Atick:1987kd}, one should introduce the so-called cut abelian differentials, which are defined as doubly-periodic holomorphic funtions on the torus yet have the appropriate monodromy around the branch points. 
By using the theta function $\vartheta_1(z)\equiv \vartheta_1(z|\tau)$, there are two sets of cut abelian differentials:
\begin{equation}
	\label{eq:abeliandifferentials}
	\begin{split}
		w_{n-k}^{\alpha_i}(z)&=\prod_{i=1}^{2m}\vartheta_1(z-z_i)^{-(1-\frac{k_i}{n})}\vartheta_1(z-z_{\alpha_i}-p_1)\prod_{j\ne i}\vartheta_1(z-z_{\alpha_j}),\\
		w_k^{\beta_i}(z)&=\prod_{i=1}^{2m}\vartheta_1(z-z_i)^{-\frac{k_i}{n}}\vartheta_1(z-z_{\beta_i}-p_2)\prod_{j\ne i}\vartheta_1(z-z_{\beta_j}),
	\end{split}
\end{equation}
where $\{z_{\alpha_i}\}$ is an arbitrary chosen set of $m$ branch points and $\{z_{\beta_i}\}$ are the rest ones.
The two shifts $p_1$ and $p_2$ are given by 
\begin{equation}\label{eq:p1p2}
	p_1=\sum_{i=1}^{2m}(1-\frac{k_i}{n})z_i-\sum_{i=1}^{m}z_{\alpha_i}, \quad p_2=\sum_{i=1}^{2m}\frac{k_i}{n}z_i-\sum_{i=1}^{m}z_{\beta_i}.
\end{equation}
Knowing the property of theta functions, one can easily check that the cut abelian differentials are doubly-periodic. 
Since $\theta_1(z-\omega)\sim (z-\omega)$ as $z\rightarrow \omega$, one can see that the cut abelian differentials also satisfy the appropriate local monodromy around each branch points.
It is also worth to mention that, within each set, the cut abelian differentials are independent. However, there is no requirement that $w_{n-k}^{\alpha_i}(z)$ should be independent of $w_{k}^{\beta_i}(z)$.

From the pole structure of $z \rightarrow \omega $, one can fix the Green's function up to the most general form
\begin{equation}
	\label{eq:greenfunctionsolution_undetermined}
	\begin{split}
		g(z,\omega)&= g_s(z,\omega)-\sum_{i=1}^{m}\sum_{j=1}^{m}A_{ij}w_k^{\beta_j}(\omega) w_{n-k}^{\alpha_i}(z),\\
		h(\bar z,\omega)&= \sum_{i=1}^{m}\sum_{j=1}^{m} B_{ij}w_{k}^i(\omega) \bar{w}_{k}^j(\bar z),
	\end{split}
\end{equation}
where $g_s(z,w)$ is singular part of the Greens function 
\begin{equation}
	\label{eq:gs}
	g_s(z,\omega)=\prod_{i=1}^{2m}\vartheta_1(z-z_i)^{-(1-\frac{k_i}{n})}\prod_{i=1}^{2m}\vartheta_1(w-z_i)^{-\frac{k_i}{n}}\left[\frac{\vartheta_1^\prime (0)}{\vartheta_1(z-\omega)}\right]^2 P(z,\omega).
\end{equation}
It can be shown that the exact form of $P(z,\omega)$ turns out to be irrelevant in the end \cite{Atick:1987kd}.

\begin{figure}
	\centering
	\includegraphics[width=0.8\textwidth]{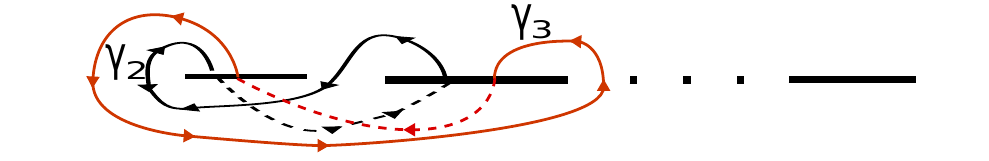}
	\caption{We can always put the twist fields at the left ends and the antitwist fields at the right ends. The branched cuts can be chosen as connecting branch points $z_{2i-1}$ to $z_{2i}$, so that the number of branch cuts is $m$. Besides the two cycles of the original torus, the remaining loops $\gamma_a,a=1,\cdots,2m-2$, which encloses branch points $z_1$ and $z_{a+1}$ respectively.}\label{fig:torus2}
\end{figure}

To determine the coefficients $A_{ij}$ and $B_{ji}$ in \eqref{eq:greenfunctionsolution_undetermined}, one can impose the global monodromy conditions:
\begin{equation}\label{eq:globalqu}
	\oint_{C_a} dz g(z,\omega)+\oint_{C_a} d \bar z h(\bar z,\omega)=0,
\end{equation}
where $C_a$ are the independent net twist zero loops on the torus. The global monodromy conditions just mean that $\phi$ is single valued around the closed loops $C_a$. There are  two cycles inherit from the original torus, while the remaining loops are chosen as shown in figure \ref{fig:torus2}.

\begin{figure}
	\centering
	\includegraphics[width=0.8\textwidth]{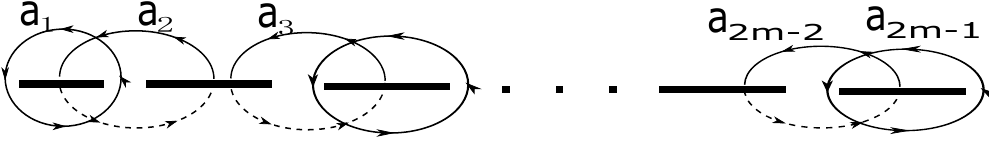}
	\caption{The solid line is on the first sheet, while the broken line is on the $n$th sheet  }\label{fig:torus}
\end{figure}

These equations \eqref{eq:globalqu} can be solved by introducing the cut period matrix
$\mathbf{W}_a^i$ defined by
\begin{equation}
	\label{eq:cutperiodmatrix}
	\begin{split}
		\mathbf{W}_a^i &\equiv \oint _{C_a} dz w_{n-k}^{\alpha_i}(z), \,i=1,\cdots,m,\\
		\mathbf{W}_a^{m+j} &\equiv \oint_{C_a} d\bar z \bar{w}_{k}^{\beta_j}( \bar z),\,j=1,\cdots, m.
	\end{split}
\end{equation}
However we have to say, the loops chosed in figure \ref{fig:torus2} are not suitable for doing numerical integration. As an alternative, we can use the contours $\alpha_a$ described in figure \ref{fig:torus}.
The relations between $\gamma_a$ and $\alpha_a$ are simple:
\begin{equation}
	\gamma_1=\alpha_1, \, \gamma_a=\alpha_a-\gamma_{a-1} \text{ for }  1<a\le 2m-2.
\end{equation}
Further, by using the OPEs \eqref{eq:ope}
one can get
\begin{equation}
	\label{eq:Tz}
	\begin{split}
		\braket{\braket{T(z;z_i)}}&\equiv \frac{\braket{T(z)\prod_{i=1}^{2m}\sigma_{k_i}(z_i,\bar{z}_i)}}{\braket{\prod_{i=1}^{2m}\sigma_{k_i}(z_i,\bar{z}_i)}}= \lim_{\omega\rightarrow z}\left [ g(z,\omega;z_i)-\frac{1}{(z-\omega)^2} \right ]\\
		\partial_{z_i}\ln Z_{qu}&=\lim_{z\rightarrow z_i}\left[ (z-z_i)\braket{\braket{T(z;z_i)}}-\frac{h_i}{z-z_i} \right],
	\end{split}
\end{equation}
where the second equation of \eqref{eq:Tz} is essentially the conformal Ward identity.
By integrating the conformal Ward identity, one can obtain the quantum part up to an unfixed normalization function $N$.
Since the middle calculation is rather straightforward but lengthy, we will not repeat them here, one can find all the missing steps in Ref.\cite{Atick:1987kd}. Finally, multiplying the quantum part for different $k$-modes together, we have
\begin{equation}
	\label{eq:quantumpart}
	\begin{split}
		Z_{qu}=&\frac{N}{(\det \Im \tau) |\eta(\tau)|^4}\left(\prod_{k=1}^{n-1}\vartheta_1(p_1)^{(m-1)(n-1)} \bar{\vartheta}_1(p_2)^{(n-1)(m-1)}\right) \prod_{i<j}^{m}\vartheta^{n-1}_{\alpha_i\alpha_j}\\
		&\quad \prod_{i<j}^{m}\bar{\vartheta}^{n-1}_{\beta_i\beta_j} \left(\prod_{k=1}^{n-1}|\det{\mathbf{W}}|^{-1}\prod_{i<j}^{2m}\vartheta_{ij}^{-(1-\frac{k_i}{n})(1-\frac{k_j}{n})}(\bar{\vartheta}_{ij})^{-\frac{k_i}{n}\frac{k_j}{n}}\right),
	\end{split}
\end{equation}
where we have denoted $\vartheta_1(z_i-z_j)$ by $\vartheta_{ij}$. Note that the factor
$\frac{1}{(\det \Im \tau) |\eta(\tau)|^4}$ is coming from the $k=0$ mode, which means that there are no twist field insertions, i.e., it is just the torus partition function of a free complex boson \cite{cft}.
For each $k$-mode, since there are pairs of twist-antitwist insertions, one can set $k_i=k$ for $i$ odd and $k_i=n-k$ for $i$ even.
Here $N$ is the normalization function came from the integration of conformal Ward identity, which can be fixed by factorizing the correlation function on to lower genus ones. 

\section{Classical partition function on $\mathcal{T}_{n,m}$}

As we mentioned earlier, in \eqref{eq:totalcorrelationfunction}, the winding numbers for different $k$-modes are correlated due to the compactification condition. Hence if one tries to compute the classical solution for different $k$-modes separately, then one has to introduce the redundant winding numbers. As a result, in the final summation, there will be some unpleasant infinity need to be regularized. In general, the regularization is quite involved. Roughly speaking, one has to find the zero eigenvectors of some complicated matrix. A concrete example of regularization can be found in \cite{Calabrese:2009ez}. However, for more general cases, this regularization process could be very complicated.
Therefore, we are not going to calculate the classical solution for different $k$-mode separately as in \cite{Calabrese:2009ez}. Instead, without using the trick of replicated target space, we construct the independent classical solutions for the $n$-sheeted covering surface directly. Such that there are no redundant summation. Actually, this similar more direct strategy has been implemented to calculate the higher-genus correlation functions of WZW models long time ago\cite{Naculich:1989ii}.

The classical action is given by
\begin{equation}
	\label{eq:classicalaction}
	S=\frac{1}{8\pi}\int_{\mathcal{T}_{n,m}}\left(\partial \phi(z)\dd z\wedge \bar{ \partial} \bar{ \phi}(\bar{z}) \dd \bar{z}+ \partial \bar{ \phi}(z)\dd z\wedge \bar{ \partial}  \phi(\bar{z}) \dd \bar{z}\right).
\end{equation}
From the classical equation of motion we know that $\partial \phi(z)\dd z$ and $\partial \bar{ \phi}(z)\dd z$ are the holomorphic one forms defined on $\mathcal{T}_{n,m}$, while $\bar{ \partial} \bar{ \phi}(\bar{z}) \dd \bar{z}$ and $\bar{ \partial}  \phi(\bar{z}) \dd \bar{z}$ are anti-holomorphic. 
Thus, if we denote the basis of holomorphic one forms by $w_{i}(z)$, then the classical solutions can be written as a linear summation of the basis
\begin{equation}
	\partial \phi(z)\dd z=\alpha^i w_i(z), \quad \partial \bar{\phi}(z)\dd z=\beta^i w_j(z).
\end{equation}
To fix the coefficients $\alpha^i$ and $\beta^i$, we need to use the global monodromy conditions similar as \eqref{eq:globalqu}, only for this time there will be a shift on the RHS due to the windings. For simplicity, we will assume both the real and imaginary part of the compactified radii are $R$, then the global monodromy condition is
\begin{equation}
	\label{eq:globalmonodromy}
	\oint_{C_a}\partial \phi(z) \dd z+\oint_{C_a}\bar{\partial}\phi(\bar{z})\dd \bar{z}= 2\pi R(m_a+\ii n_a)\equiv v_a,\quad m_a,n_a \in \mathbb{Z}.
\end{equation}
It is important to notice that the number of independent closed loops is the same as 
the number of independent holomorphic one forms \cite{Atick:1987kd}.
Plugging the solution of $\alpha^i$ and $\beta^i$ back into \eqref{eq:classicalaction}, the classical action can always be written as
\begin{equation}
	\label{eq:classical}
	\begin{split}
		S_{cl} = \frac{1}{8\pi} v_a M^{ab}\bar v_b.
	\end{split}
\end{equation}
Here $M^{ab}$ is dependent on the choice of homology basis, although $S_{cl}$ doesn't. Thus, one can always find a convenient basis to simplify $M^{ab}$. The procedure is explained as follows.

We first recall some basic definitions.
Let $a_i, b_i, i = 1,\cdots,g$, be a canonical homology basis of genus $g$ Riemann surface $\mathcal{R}$
and let $w_k , k = 1,\cdots,g$, be the dual basis of $H^1(\mathcal{R},\mathbb{C})$.
Then the period matrix of $\mathcal{R}$ is defined by
\begin{equation}
	\mathbf{P}=(\mathbf{A},\mathbf{B}),
\end{equation}
where
\begin{equation}
	\mathbf{A}_{ij}=\oint_{a_i}w_j, \quad \mathbf{B}_{ij}=\oint_{b_i}w_j.
\end{equation}
For many cases, it is more convenient to use the canonical basis of $H^1(\mathcal{R},\mathbb{C})$ defined  by
\begin{equation}
	\tilde{w}_k=\sum_i(\mathbf{A}^{-1})_{ki}w_i.
\end{equation}
Then the Riemann matrix of the surface $\mathcal{R}$ can be defined by
\begin{equation}
	\mathbf{\Omega}_{ij}=\oint_{b_i}\tilde{w}_j=(\mathbf{B} \cdot \mathbf{A}^{-1} )_{ij}.
\end{equation}
By using the Riemann bilinear relation \cite{Bobenko:2011} 
\begin{equation}
	\begin{split}
		(w_i,w_j)&\equiv \ii \int_{\mathcal{R}}w_i \wedge \bar{w_j}=\ii \sum_{k=1}^g \oint_{a_k}w_i \oint_{b_k}\bar{w_j}-\oint_{b_k}w_i \oint_{a_k}\bar{w_j}
	\end{split}
\end{equation}
and arranging the loops $\{C_1,\cdots,C_{2g}\}$ to be $\{a_1,\cdots,a_g,b_1,\cdots,b_g\}$, we have
\begin{equation}
	M^{ab}=\left[(\mathbf{G}^{-1})^{\mathrm{T}}\cdot \mathbf{H} \cdot (\bar{\mathbf{G}}^{-1}) \right]^{ab},
\end{equation}
where
\begin{equation}
	\mathbf{G}=\begin{bmatrix}
		\mathbf{1}_g & \mathbf{1}_g\\
		\mathbf{\Omega} & \bar{\mathbf{\Omega}}
	\end{bmatrix},
	\mathbf{H}=
	\begin{bmatrix}
		2\ii \Im \mathbf{\Omega}& 0\\
		0 & 2\ii \Im \mathbf{\Omega}
	\end{bmatrix}.
\end{equation}
A straightforward block-matrix calculation shows that 
\begin{equation}
	\label{eq:m}
	M=
	\left(\begin{array}{c|c}
		\Im{\Omega}+\Re{\Omega}\cdot \Im{\Omega}^{-1} \cdot \Re{\Omega}\quad &\quad\Re{\Omega}\cdot \Im{\Omega}^{-1}\\
		\hline
		\Im{\Omega}^{-1}\cdot \Re{\Omega} & \quad \Im{\Omega}^{-1}
	\end{array}\right).
\end{equation}
Putting all things together, the classical part of the partition function can be written as
\begin{equation}\label{eq:zcl}
	\begin{split}
		\sum_{\mathrm{windings}}\exp(-S_{cl})&=\sum_{\mathrm{\{m_a,n_a\}}}\exp\left[\frac{-1}{8\pi}v_a M^{ab}\bar{v}_b\right]=\left | \Theta(0|\ii \frac{R^2}{2}M)\right |^2.
	\end{split}
\end{equation} 
Here function $\Theta(0|\Omega)$ is the multi-dimensional theta function \cite{nist}.
We have noticed that the form of the classical summation \eqref{eq:zcl} is the same as \cite{Coser:2013qda}.
Thus the classical partition function should only depend on the Riemann matrix of the surface. However, it is very hard to evaluate the Riemann matrix for a generic Riemann surface. The most well studied examples are the plane algebraic curves, i.e., the branched covering of $\mathbb{CP}^1$, for which one can find the Riemann matrix numerically \cite{Bobenko:2011,Coser:2013qda}. For our case $\mathcal{T}_{n,m}$, which is defined as the $n$-sheeted covering of the torus gluing along $m$ branch cuts, the problem is quite different. Luckily, by using the powerful theta functions, we are able to propose a basis of $H_1(\mathcal{T}_{n,m},\mathbb{Z})$ and its dual basis of $H^1(\mathcal{T}_{n,m},\mathbb{C})$, such that the explicit form of the classical partition function can be found in terms of theta functions.

In the following we will construct the basis of $H^1(\mathcal{T}_{n,m},\mathbb{C})$. Notice that the genus of $\mathcal{T}_{n,m}$ is given by
$
g=n+(m-1)(n-1)=m(n-1)+1,
$
which also gives the number of independent holomorphic one-forms.
Remember that there is only one nontrivial holomorphic one-form for the torus, i.e., $\dd z$. So all we need to do is to construct the rest $m(n-1)$ differentials, which should encode the information of the gluing procedure. 
Let's now look more closely into the cut abelian differentials defined in \eqref{eq:abeliandifferentials}, one should notice that they are nothing but a subsets of $H^1(\mathcal{T}_{n,m},\mathbb{C})$. Actually, there is another example: For free boson on the singular $Z_n$ curve \eqref{eq:zncurve}, the collection of cut abelian differentials for $k=1,\cdots,n-1$ are exactly the basis of the holomorphic one-forms of curve \eqref{eq:zncurve} \cite{Dixon:1986qv,Bershadsky:1986fv}. Hence, one can similarly construct the basis of $H^1(\mathcal{T}_{n,m},\mathbb{C})$ out of the cut abelian differentials.

Inspired by the work of \cite{Atick:1987kd}, we denote the branch points by $\{z_1,z_2,\cdots, z_{2m}\}$, so that there are $m$ branch cuts connecting $z_{2i-1}$ to $z_{2i}$ for $i=\{1,\cdots,m\}$.
We propose the basis of holomorphic one-forms to be
\begin{equation} \label{eq:holomorphicdifferentials}
	\begin{split}
		&w_{s,i}(z)\dd z=w_{i+(s-1)m}(z)\dd z=\gamma_s(z)\vartheta_1(z-z_{2i-1}-Y_s)\prod_{j\ne i}^m\vartheta_1(z-z_{2j-1})\dd z,\\
		& s\in \{1,\cdots,n-1\}, i \in \{1,\cdots,m\}, \quad w_{n,1}\dd z=w_{g}\dd z=\dd z
	\end{split}
\end{equation}
where
\begin{equation}
	\begin{split}
		\gamma_s(z)&\equiv \prod_{i=1}^{m}\vartheta_1(z-z_{2i-1})^{-(1-\frac{s}{n})}\vartheta_1(z-z_{2i})^{-\frac{s}{n}},\\
		Y_s&=\left( \{ \frac{s(n-1)}{n}\}-1\right)\sum_{i=1}^{m}z_{2i-1}+\frac{s}{n}\sum_{i=1}^{m}z_{2i}.
	\end{split}
\end{equation}
Here $\{\bullet\}$ is a convention representing the fractional part of $\bullet$.
Since $w_{s,i}(z)\dd z$ are just the independent cut abelian differentials for the $k=s$ mode defined in \cite{Atick:1987kd}, as a consequence, the basis in \eqref{eq:holomorphicdifferentials} are all independent by construction. One can also see that, the number of independent holomorphic one-forms is exactly the genus $g$.

\begin{figure}[h]
	\centering
	\includegraphics[width=0.7\textwidth]{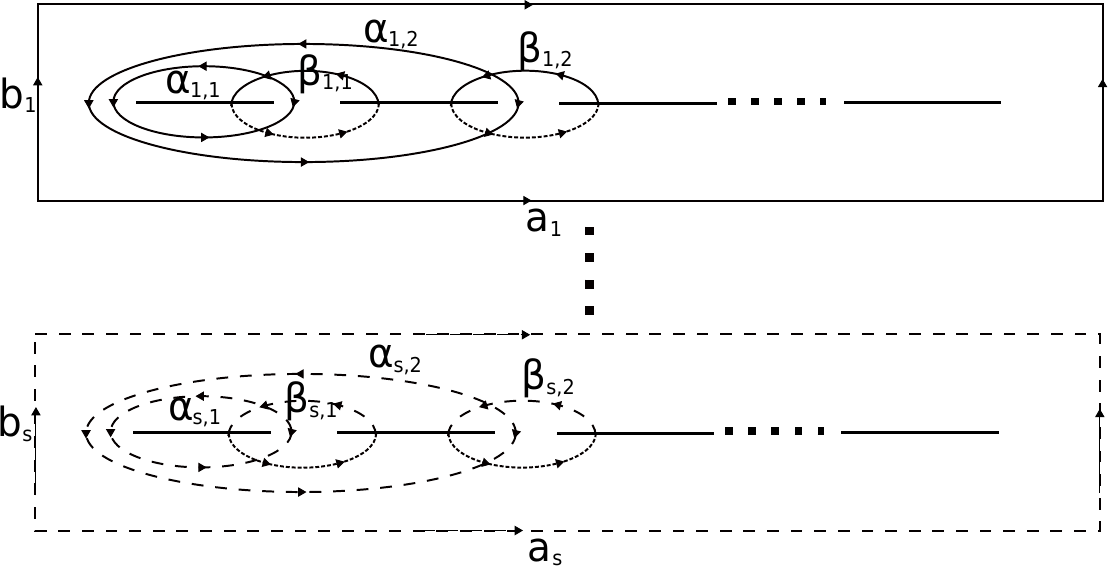}
	\caption{The canonical homology basis of $\mathcal{T}_{n,m}$, where $a_s$, $b_s$ are the canonical cycles of the $s$-th sheet torus. The $\alpha$ and $\beta$ cycles are chosen similarly as the algebraic curve in \cite{enolski:2006}. The solid line is on the first sheet, the broken line is on the $s$-th sheet, while the dotted line is on the $n$-th sheet. In each sheet, the number of $\alpha$ and $\beta$ cycles are both $m-1$.}\label{fig:branchtorus}
\end{figure}

Following \cite{enolski:2006}, we are going to choose a basis of the homology a-cycles $\alpha_a$ and b-cycles $\beta_a$ similar as curve \eqref{eq:zncurve}, except that there are additional loops inherited from the torus. As shown in figure \ref{fig:branchtorus}, they are denoted by
\begin{equation}\label{eq:canonicalbasis}
	\begin{split}
		&\alpha_{s,i}=\alpha_{(s-1)(m-1)+i}, \,\beta_{s,i}=\alpha_{(s-1)(m-1)+i},\\
		&s\in \{1,\cdots,n-1\}, \, i \in \{1,\cdots,m-1\},\\
		&a_{l}=\alpha_{(n-1)(m-1)+l}, \, b_{l}=\beta_{(n-1)(m-1)+l},\quad l\in \{1,\cdots, n\}.
	\end{split}
\end{equation}
We think the subscript $\{s,i\}$ is useful for the bookkeeping: $s$ label the sheet, counting from top to bottom, while $i$ label the branch points of odd order, counting from left to right.

In order to do numerical calculation, the integrals in the period matrix should be performed on the first sheet. Let's define the $\mathbf{A}$ and $\mathbf{B}$ periods as
\begin{equation}\label{eq:4integral}
	\begin{split}
		&\mathbf{A}_{s,r}^{i,j}=\oint_{\alpha_{s,i}}w_{r,j}, \, \mathbf{B}_{s,r}^{i,j}=\oint_{\beta_{s,i}}w_{r,j} \\
		&\mathbf{A}_{l,r}^{j}=\oint_{a_{l}}w_{r,j},\, \mathbf{B}_{l,r}^{j}=\oint_{b_{l}}w_{r,j}\\
		&i=1,\cdots,m-1,\,j=1,\cdots,m\\
		&l=1,\cdots,n, \, r=1,\cdots,n-1
	\end{split}
\end{equation}
The contour integral of $\dd z$ is trivial, the only non-zero integrals are
\begin{equation}
	\oint_{a_l}\dd z=1, \,\oint_{b_l}\dd z=\tau,
\end{equation}
where $\tau$ is the moduli of the torus.
The integrals in \eqref{eq:4integral} are related to the integral on the first sheet by a phase
\begin{equation}
	\begin{split} \label{eq:first3}
		\mathbf{A}_{s,r}^{i,j}=\rho^{r(s-1)}\mathbf{A}_{1,r}^{i,j},\,\mathbf{A}_{l,r}^{j}=\rho^{l-1}\mathbf{A}_{1,r}^{j}, \,\mathbf{B}_{l,r}^{j}=\rho^{l-1}\mathbf{B}_{1,r}^{j},
	\end{split}
\end{equation}
where $\rho\equiv e^{2\pi \ii /n}$.
For $\mathbf{B}_{s,r}^{i,j}$, things get a little bit tricky, but we notice that
\begin{equation}
	\mathbf{B}_{s,r}^{i,j}=(\mathbf{B}_{s,r}^{i,j}-\mathbf{B}_{s+1,r}^{i,j})+(\mathbf{B}_{s+1,r}^{i,j}-\mathbf{B}_{s+2,r}^{i,j})+\cdots (\mathbf{B}_{n-2,r}^{i,j}-\mathbf{B}_{n-1,r}^{i,j})+\mathbf{B}_{n-1,r}^{i,j}.
\end{equation}
Further, the action of the cyclical automorphism $J$ on the homology basis tells us that \cite{enolski:2006}
\begin{equation}
	\begin{split}
		&J(\beta_{s,j})=\beta_{s+1,j}-\beta_{1,j}, \, s=1,\cdots,n-2,\\
		&J(\beta_{s+1,j}-\beta_{s,j})=\beta_{s+2,j}-\beta_{s+1,j}, \quad J(\beta_{n-1,j})=-\beta_{1,j}
	\end{split}
\end{equation}
from which one can see that
\begin{equation}
	\rho^r(\mathbf{B}_{s,r}^{i,j}-\mathbf{B}_{s+1,r}^{i,j})=(\mathbf{B}_{s+1,r}^{i,j}-\mathbf{B}_{s+2,r}^{i,j}).
\end{equation}
Then we have
\begin{equation}\label{eq:last1}
	\begin{split}
		\mathbf{B}_{s,r}^{i,j}&=\rho^{r(s-1)}(\mathbf{B}_{1,r}^{i,j}-\mathbf{B}_{2,r}^{i,j})+ \rho^{rs}(\mathbf{B}_{1,r}^{i,j}-\mathbf{B}_{2,r}^{i,j})+\cdots+\rho^{r(n-2)}(\mathbf{B}_{1,r}^{i,j}-\mathbf{B}_{2,r}^{i,j})\\
		&=\frac{\rho^{r(s-1)}(1-\rho^{r(n-s)})}{1-\rho^r}(\mathbf{B}_{1,r}^{i,j}-\mathbf{B}_{2,r}^{i,j})\equiv \frac{\rho^{r(s-1)}(1-\rho^{r(n-s)})}{1-\rho^r}\mathbf{C}_{1,r}^{i,j},
	\end{split}
\end{equation}
where we have defined $\mathbf{C}_{1,r}^{i,j}=\mathbf{B}_{1,r}^{i,j}-\mathbf{B}_{2,r}^{i,j}$, which is an integral along a contour just enclosing the branch points $z_{2i}$ and $z_{2i+1}$. 

Now, from \eqref{eq:first3} and \eqref{eq:last1}, one can see that all the integrals in the period matrix can written as the integrals on the first sheet, which can be calculated by contour integrals:
\begin{equation}\label{eq:firstsheetintegral}
	\begin{split}
		\mathbf{A}_{1,r}^{i,j}&=(\rho^{-r}-1)\sum_{k=1}^{i}\int_{z_{2k-1}}^{z_{2k}} w_{r,j},\\
		\mathbf{C}_{1,r}^{i,j}&=\rho^{r/2}(\rho^{-r}-1)(-1)^{1/n}\int_{z_{2i}}^{z_{2i+1}} w_{r,j},\\
		\mathbf{A}_{1,r}^j&=\int_{0}^{1}w_{r,j},\quad \mathbf{B}_{1,r}^j=\int_{0}^{\ii \beta} w_{r,j}.
	\end{split}
\end{equation}
By now, all the elements in the period matrix $(\mathbf{A},\mathbf{B})$ can be numerically computed, so is the Riemann matrix $\mathbf{\Omega}$.
To be more concrete, we give some examples of calculation in the appendix.

Finally,  we are able to give the most general form of the partition function on $\mathcal{T}_{n,m}$:
\begin{equation}
	\begin{split} \label{eq:torus-general}
		Z_{\mathcal{T}_{m,n}}=&\frac{N}{(\det \Im \tau) |\eta(\tau)|^4}\left(\prod_{k=1}^{n-1}|\vartheta_1(p)|^{2(m-1)(n-1)} \right) \prod_{i<j}^{m}\vartheta^{n-1}_{\alpha_i\alpha_j}\prod_{i<j}^{m}\bar{\vartheta}^{n-1}_{\beta_i\beta_j}\\
		& \left(\prod_{k=1}^{n-1}|\det{\mathbf{W}}|^{-1}\prod_{i<j}^{2m}\vartheta_{ij}^{-(1-\frac{k_i}{n})(1-\frac{k_j}{n})}(\bar{\vartheta}_{ij})^{-\frac{k_i}{n}\frac{k_j}{n}}\right)
		\left | \Theta(0|\ii \frac{R^2}{2}M)\right |^2,
	\end{split}
\end{equation}
where $k_i=k$ for $i$ odd and $k_i=n-k$ for $i$ even and
$
p=\frac{k}{n}\sum_{i=1}^{m}(z_{2i}-z_{2i-1}).
$
If one set $\tau=\ii \beta$ to be pure imaginary and all the branch cuts lie on the real interval: $0<z_1<z_2\cdots<z_{2m-1}<z_{2m}<1$, then the partition function \eqref{eq:torus-general} corresponds to the R\'enyi entanglement entropy for $m$ disjoint intervals in a finite system at finite temperature $1/\beta$.

\section{Conclusions}

In conclusion, we obtain the partition function of free compact boson on  the $n$-sheeted covering of a torus gluing along $m$ branch cuts. In order to achieve that goal, we proposed a canonical homology basis and construct the holomorphic differentials in terms of theta functions, such that the period matrix can be directly constructed, as importantly, it is numerically computable. Given the period matrix, we are able to generalized the earlier results of $g=3$ results \cite{Liu:2015iia} to higher genus, which means that we can calculate the $n$-th R\'enyi entanglement entropy for arbitrary disjoint intervals in a finite system at finite temperature. 

Mathematically, it is also very interesting to know if there are some Thomae type formulas for branched covering of torus, which may relate the positions of the branch points on the torus to the period matrix by theta functions. The basis \eqref{eq:holomorphicdifferentials} and \eqref{eq:canonicalbasis} we proposed here  may be useful for that purpose as well. Also, we should mention that in Ref.\cite{Matone:2001uy}, there is a theorem (Theorem 3.) which states that the period matrix of the branch covering of the torus should satisfy the conditon:
\begin{equation}
	m_j^\prime -\sum_{k=1}^{g}\Omega_{ij}n_k^\prime=\bar{c}(m_j^\prime,n_k^\prime;m_j,n_k) \left(m_j-\sum_{k=1}^{g}\Omega_{ij}n_k \right), \quad m_j^\prime,n_k^\prime,m_j,n_k \in \mathbb{Z}.
\end{equation}
One may use the method we proposed to calculate the coefficient $\bar{c}(m_j^\prime,n_k^\prime;m_j,n_k)$.

\section{Acknowledgement}
We would like to thank Tianjun Li for carefully reading the manuscript. We also want to thank Wei Fu, Lina Wu for useful discussing.

\appendix

\section{Example: $\mathcal{T}_{2,2}$}

We are going to discuss $2$-sheeted covering of torus with four branch points, which was studied in \cite{Liu:2015iia}. The basis of holomorphic differentials are given by
\begin{equation}
	\begin{split}
		&w_{1,1}=\gamma(z)\vartheta_1(z-z_1-Y_1)\vartheta_1(z-z_3)\dd z\\
		&w_{1,2}=\gamma(z)\vartheta_1(z-z_3-Y_1)\vartheta_1(z-z_1)\dd z, \quad w_3=\dd z,
	\end{split}
\end{equation}
where
\begin{equation}
	Y_1=\frac{1}{2}(z_4-z_3+z_2-z_1),\, \gamma(z)=\prod_{i=1}^{4}\vartheta_1(z-z_i)^{\frac{1}{2}}.
\end{equation}
The $\mathbf{A}$ and $\mathbf{B}$ periods are given in terms of the cut period matrix \eqref{eq:cutperiodmatrix}
\begin{equation}
	\mathbf{A}=
	\begin{pmatrix}
		{W_4}^1 & {W_4}^2 & 0 \\
		{W_1}^1 & {W_1}^2 & 1 \\
		-{W_1}^1 & -{W_1}^2 & 1
	\end{pmatrix},
	\,\mathbf{B}=
	\begin{pmatrix}
		{W_2}^1 & {W_2}^1 & 0 \\
		-{W_3}^1 & -{W_3}^1 & \tau \\
		{W_3}^1 & {W_3}^2 & \tau
	\end{pmatrix}.
\end{equation}
They are the contour integrals on the first sheet defined by \eqref{eq:firstsheetintegral}:
\begin{equation}
	\begin{split}
		W_2^{1(2)}& =\oint_{\gamma_2} \dd z \omega^1(z)=(e^{2 \pi \mathrm{i}\frac{1}{2}}-1)\int_{z_2}^{z_3}\dd z w^{1(2)}(z)\\
		{W_3}^{1(2)} & =\oint_{\gamma_3} \dd z w^{1(2)}(z)=\int_{0}^{-\tau }w^{1(2)}\dd z \\
		{W_4}^1 &=e^{-\frac{\mathrm{i} \pi}{2}}2 \mathrm{i} \sin(\frac{3}{2}\pi)(-1)^{-1/2}\int_{z_1}^{z_2}\dd z \vartheta_1(z_1-z)^{-1/2}\vartheta_1(z-z_2)^{1/2} \vartheta_1(z-z_3)^{1/2}\vartheta_1(z-z_4)^{-1/2}\\
		{W_4}^2 &=e^{-\frac{\mathrm{i} \pi}{2}}2 \mathrm{i} \sin(\frac{3}{2}\pi)(-1)^{1/2}\int_{z_1}^{z_2}\dd z \vartheta_1(z_1-z)^{1/2}\vartheta_1(z-z_2)^{-1/2} \vartheta_1(z-z_3)^{-1/2}\vartheta_1(z-z_4)^{1/2}.
	\end{split}
\end{equation}
In order to simplify the formula, we define
\begin{equation}
	\begin{split}
		x&= \mathrm{i} \frac{-{W_2}^2{W_3}^1+{W_2}^1{W_3}^2}{{W_1}^2{W_2}^1-{W_1}^1{W_2}^2}\\
		y&= \mathrm{i} \frac{-2{W_1}^2{W_3}^1+2{W_1}^1{W_3}^2+{W_2}^2{W_4}^1-{W_2}^1{W_4}^2}{-4{W_1}^2{W_2}^1+4{W_1}^1{W_2}^2}\\
		z&= \mathrm{i} \frac{-{W_1}^2{W_4}^1+{W_1}^1{W_4}^2}{-2{W_1}^2{W_2}^1+2{W_1}^1{W_2}^2},
	\end{split}
\end{equation}
and also assuming that $\tau$ is pure imaginary $\tau=\ii \beta$, then the period matrix $\Omega=\mathbf{B}\cdot\mathbf{A}^{-1}$ have a simpler form:
\begin{equation}\label{eq:omega22}
	\Omega=
	\left(
	\begin{array}{ccc}
		\frac{\ii}{2 z} & \,-\frac{y}{2 z} &\, \frac{y}{2 z} \\
		-\frac{y}{2 z} & \,\ii \left(\frac{x z-y^2}{2 z}+\frac{\beta }{2}\right) &\, \ii \left(\frac{y^2-x z}{2 z}+\frac{\beta }{2}\right) \\
		\frac{y}{2 z} & \, \ii \left(\frac{y^2-x z}{2 z}+\frac{\beta }{2}\right) & \,\ii \left(\frac{x z-y^2}{2 z}+\frac{\beta }{2}\right) \\
	\end{array}
	\right).
\end{equation}
where we have used a rather nontrivial results in \cite{Liu:2015iia}
\begin{equation}\label{eq:mir}
	y^2-xz=\frac{1}{2}\frac{{W_3}^1{W_4}^2-{W_4}^1{W_3}^2}{{W_1}^1{W_2}^2-{W_2}^1{W_1}^2}.
\end{equation}
The equality \eqref{eq:mir} actually ensures the $T$-duality of the classical action \cite{Liu:2015iia}.
One can also numerically check that \eqref{eq:omega22} is indeed a Riemann matrix, i.e., $\Im{\Omega}$ is symmetric and positive definite.

\end{document}